\documentclass[reprint,amssymb,amsmath,superscriptaddress,aps,showpacs,10pt,floatfix,longbibliography,prl]{revtex4-1}

\usepackage[pdftex]{graphicx} 
\usepackage{epstopdf}
\usepackage{verbatim}
\usepackage{amsmath}
\usepackage{color}
\usepackage{subfigure}
\usepackage{amsbsy}
\usepackage{wasysym}

\begin{document}

\title{Rearrangement of uncorrelated valence bonds\\ evidenced by low-energy spin excitations in YbMgGaO$_4$}

\author{Yuesheng Li}
\email{yuesheng.man.li@gmail.com}
\affiliation{Experimental Physics VI, Center for Electronic Correlations and Magnetism, University of Augsburg, 86159 Augsburg, Germany}

\author{Sebastian Bachus}
\affiliation{Experimental Physics VI, Center for Electronic Correlations and Magnetism, University of Augsburg, 86159 Augsburg, Germany}

\author{Benqiong Liu}
\affiliation{Key Laboratory of Neutron Physics, Institute of Nuclear Physics and Chemistry, CAEP, Mianyang 621900, PR China}
\affiliation{J\"{u}lich Centre for Neutron Science (JCNS) at Heinz Maier-Leibnitz Zentrum (MLZ), Forschungszentrum J\"{u}lich GmbH, Lichtenbergstrasse 1, 85748 Garching, Germany}

\author{Igor Radelytskyi}
\affiliation{J\"{u}lich Centre for Neutron Science (JCNS) at Heinz Maier-Leibnitz Zentrum (MLZ), Forschungszentrum J\"{u}lich GmbH, Lichtenbergstrasse 1, 85748 Garching, Germany}

\author{Alexandre Bertin}
\affiliation{Institut fuer Festkoerperphysik, TU Dresden, D-01062, Dresden, Germany}

\author{Astrid Schneidewind}
\affiliation{J\"{u}lich Centre for Neutron Science (JCNS) at Heinz Maier-Leibnitz Zentrum (MLZ), Forschungszentrum J\"{u}lich GmbH, Lichtenbergstrasse 1, 85748 Garching, Germany}

\author{Yoshifumi Tokiwa}
\affiliation{Experimental Physics VI, Center for Electronic Correlations and Magnetism, University of Augsburg, 86159 Augsburg, Germany}

\author{Alexander A. Tsirlin}
\affiliation{Experimental Physics VI, Center for Electronic Correlations and Magnetism, University of Augsburg, 86159 Augsburg, Germany}

\author{Philipp Gegenwart}
\affiliation{Experimental Physics VI, Center for Electronic Correlations and Magnetism, University of Augsburg, 86159 Augsburg, Germany}

\begin{abstract}
dc-magnetization data measured down to 40 mK speak against conventional freezing and reinstate YbMgGaO$_4$ as a triangular spin-liquid candidate. Magnetic susceptibility measured parallel and perpendicular to the $c$-axis reaches constant values below 0.1 and 0.2 K, respectively, thus indicating the presence of gapless low-energy spin excitations. We elucidate their nature in the triple-axis inelastic neutron scattering experiment that pinpoints the low-energy ($E$ $\leq$ $J_0$ $\sim$ 0.2 meV) part of the excitation continuum present at low temperatures ($T$ $<$ $J_0$/$k_B$), but \emph{completely} disappearing upon warming the system above $T$ $\gg$ $J_0$/$k_B$. In contrast to the high-energy part at $E$ $>$ $J_0$ that is rooted in the breaking of nearest-neighbor valence bonds and persists to temperatures well above $J_0$/$k_B$, the low-energy one originates from the rearrangement of the valence bonds and thus from the propagation of unpaired spins. We further extend this picture to herbertsmithite, the spin-liquid candidate on the kagome lattice, and argue that such a hierarchy of magnetic excitations may be a universal feature of quantum spin liquids.
\end{abstract}

\maketitle

\emph{Introduction.}---Quantum spin liquids (QSLs) have a special place in condensed-matter physics as states with unconventional excitations solely driven by spin degrees of freedom in the absence of charge and orbital fluctuations. The QSL physics may be behind many intriguing phenomena studied over the last decades, including the high-temperature superconductivity~\cite{anderson1973resonating,anderson1987resonating}. Exotic properties of the QSLs are also central to new technologies, such as topological quantum computing~\cite{nayak2008non}. The prototype of a QSL was proposed by Anderson back in 1973 as a resonating-valence-bond (RVB) state, a superposition of many different partitions of the triangular spin network into valence bonds (VBs, spin-0 singlets), $\frac{1}{\sqrt{2}}$($\mid\uparrow\downarrow\rangle$-$\mid\downarrow\uparrow\rangle$)~\cite{anderson1973resonating,anderson1987resonating}. In a two-dimensional QSL, unpaired spins constitute fermionic excitations with exotic properties, the quantum number fractionalization and intrinsic topological order~\cite{wen2004quantum,moessner2006geometrical,lee2008end,balents2010spin}. They propagate through the lattice by locally rearranging the uncorrelated VBs, an effect well established theoretically but never observed in any real material.

YbMgGaO$_4$, the triangular antiferromagnet proposed by one of us in 2015~\cite{li2015gapless}, may be a window into this interesting physics. It features a triangular lattice of the Yb$^{3+}$ ions with the antiferromagnetic coupling of $J_0$ $\sim$ 0.2 meV~\footnote{Refs.~\cite{li2015gapless},~\cite{li2015rare}, and~\cite{paddison2016continuous} report $J_0$ $\sim$ 0.24, 0.13, and 0.20 meV, respectively. Therefore, we use the median value of $J_0$ $\sim$ 0.2 meV throughout the Letter.}, which is equivalent to the temperature of
$\sim$ 2 K~\cite{li2015gapless,li2015rare}. No signatures of spin freezing could be seen in the heat capacity~\cite{li2015gapless}, thermal conductivity~\cite{PhysRevLett.117.267202}, and muon spin relaxation ($\mu$SR)~\cite{PhysRevLett.117.097201} measurements down to 48 mK. Although zero residual entropy ($S_m$) has been reported based on the magnetic heat capacity data down to 60 mK~\cite{li2015gapless}, the zero-field $\mu$SR rate only slightly increases from $\sim$ 0.2 $\mu$s$^{-1}$ at 50 K to $\sim$ 0.3 $\mu$s$^{-1}$ below $\sim$ 0.4 K suggesting very strong quantum fluctuations~\cite{PhysRevLett.117.097201}. A broad excitation continuum can be interpreted within the framework of a \emph{spinon} Fermi surface~\cite{shen2016spinon}, although several experimental observations have challenged this optimistic scenario. In particular, absent magnetic contribution to the thermal conductivity~\cite{PhysRevLett.117.267202} implies the localization of fermionic excitations, possibly caused by the random distribution of Mg$^{2+}$ and Ga$^{3+}$ that has strong impact on the crystal-field levels of Yb$^{3+}$~\cite{PhysRevLett.118.107202,paddison2016continuous}, although its eventual effect on the magnetic parameters and spin-liquid physics remains debated~\cite{zhang2017hierarchy}. Signatures of spin freezing observed in the ac-susceptibility around 0.1 K~\cite{PhysRevLett.120.087201} seem to corroborate the importance of structural disorder, as do some of the recent theory studies suggesting the possibility of the spin-liquid mimicry~\cite{PhysRevLett.119.157201} or the formation of a VB glass~\cite{kimchi2017valence}.

In this Letter, we critically test these scenarios by probing the low-temperature magnetization and low-energy spin excitations of YbMgGaO$_4$. We argue that the slowing down of spin fluctuations at 0.1 K affects only few spin degrees of freedom while having no serious influence on spin dynamics. This unusual dynamics can be well understood within the uncorrelated VB formalism. The high-energy excitations at $E>J_0$ are due to the breaking of nearest-neighbor (NN) VBs~\cite{li2017nearest}, whereas the low-energy ones at $E\leq J_0$ arise from the re-arrangement of VBs and propagation of unpaired spins. This renders YbMgGaO$_4$ different from the VB glass proposed in Ref.~\cite{kimchi2017valence} and can't be well captured by any theoretical model reported to date. Interestingly, similar formalism can be applied to another QSL candidate, herbertsmithite~\cite{han2012fractionalized}. We then argue that such low-energy excitations due to the re-arrangement of VBs may be generic for QSLs.

\emph{Experimental techniques.}---Large-size high-quality single crystals of YbMgGaO$_4$ ($\sim$ 1 cm) were grown by the
floating zone technique~\cite{li2015rare}. Two properly sized (10$-$20 mg) single crystals were selected for the magnetization measurements using the Faraday force magnetometer~\cite{sakakibara1994faraday,li2018absence} down to 40 mK and up to 1 T applied both parallel and perpendicular to the $c$-axis, respectively. dc and ac magnetization above 1.8 K were measured in a magnetic property measurement system (MPMS, Quantum Design) using single crystals of $\sim$ 60 mg. Eleven best-quality single crystals (total mass of $\sim$ 10 g) were selected for the cold triple-axis inelastic neutron scattering (INS) measurements on PANDA at the Heinz Maier-Leibnitz Zentrum (MLZ)~\cite{schneidewind2015panda,supple}. The international system of units is used throughout this Letter.

\begin{figure}[t]
\begin{center}
\includegraphics[width=8.7cm,angle=0]{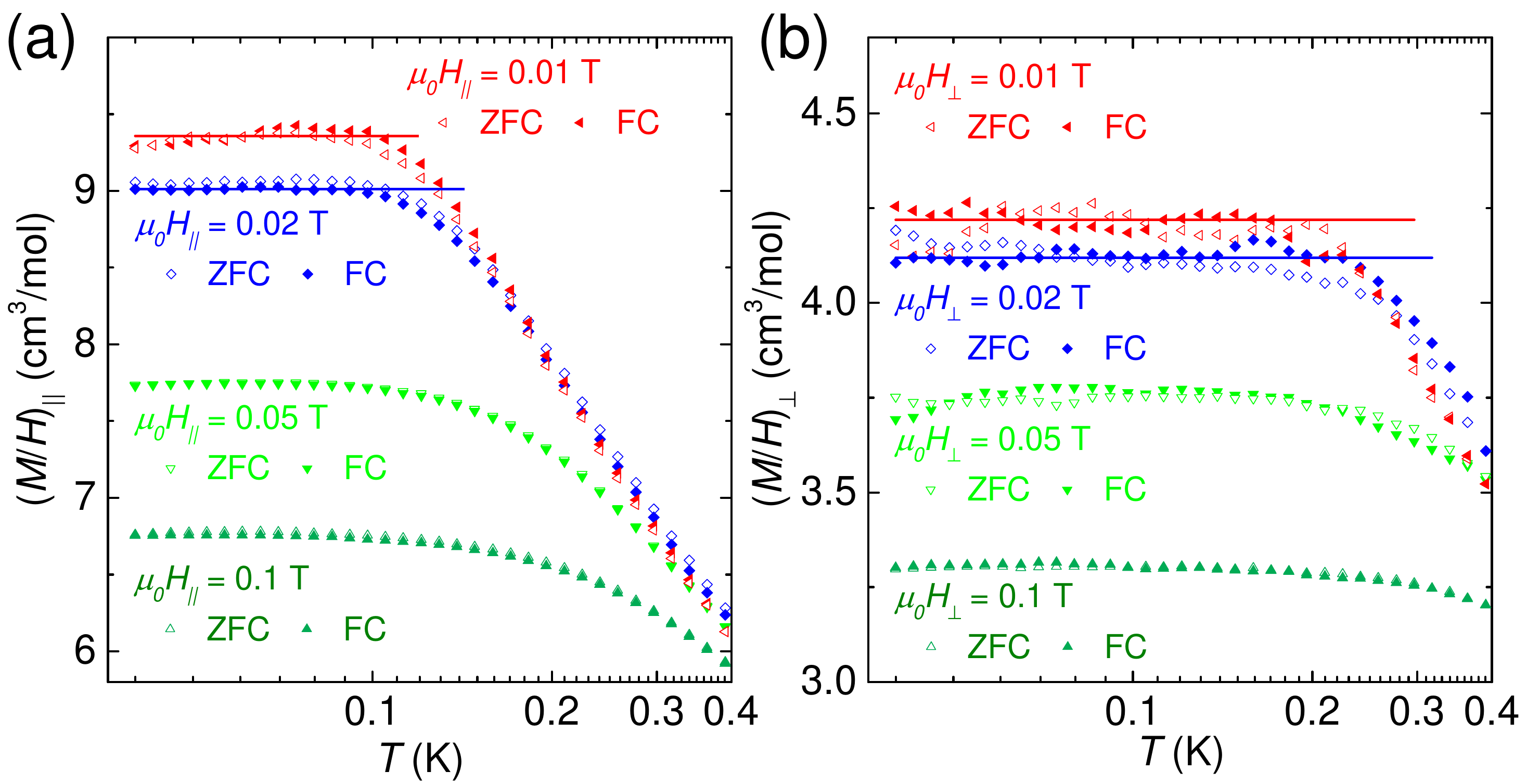}
\caption{(Color online)
Low-$T$ magnetization data of YbMgGaO$_4$ measured at selected fields applied (a) parallel and (b) perpendicular to the $c$-axis. The red and blue lines show the zero-temperature limits for the 0.01 and 0.02 T data, respectively.}
\label{fig1}
\end{center}
\end{figure}

\emph{Absence of a conventional spin freezing.}---Whereas the absence of long-range magnetic order in YbMgGaO$_4$ is well established, it remains ambiguous whether spins are static (frozen in a conventional spin glass) or dynamic at low temperatures.

Our dc magnetization measurements speak against the conventional freezing scenario. In fact, we do observe a weak anomaly (kink) at 0.1$-$0.2 K depending on the field direction. This anomaly could be paralleled to the broad peak in the ac-susceptibility at 0.1 K reported earlier~\cite{PhysRevLett.120.087201}, but several additional observations suggest that the low-temperature behavior of YbMgGaO$_4$ is different from simple freezing into a spin glass, where all or at least most of the spins would be static.

(\romannumeral1) In low fields of 0.01 and 0.02 T, the anomaly appears at $T_s^{\parallel}\sim$ 0.1 K when the field is applied along the $c$-axis [see Fig.~\ref{fig1}(a)] and at $T_s^{\perp}$ $\sim$ 0.2 K $\sim$ 2$T_s^{\parallel}$ when the field is applied in the $ab$-plane [see Fig.~\ref{fig1}(b)]. Such a direction dependence is not expected in a conventional freezing scenario, where the transition temperature should remain the same for all field directions~\cite{PhysRevB.96.094414,PhysRevB.97.054426}, especially in the applied field as low as 0.01 T, two orders of magnitude smaller than $J_0$ [$J_0/(g\mu_B)$ $\sim$ 1 T]. We recognize that $T_s^{\perp}$/$T_s^{\parallel}$ $\sim$ $J_{xx}$/$J_{zz}$, where $J_{xx}$ = 2$J_{\pm}$ $\sim$ 2 K and $J_{zz}$ $\sim$ 1 K~\cite{li2015rare},
and the direction dependence of $T_s$ merely reflects the anisotropy of magnetic couplings, whereas the difference between $\chi_{\parallel}$ and $\chi_{\perp}$ is mostly due to the $g$-tensor anisotropy~\cite{supple}. (\romannumeral2) No splitting between the ZFC and FC data is observed down to 40 mK, $|$$\chi^{ZFC}$-$\chi^{FC}$$|$/($\chi^{ZFC}$+$\chi^{FC}$) $<$ 2\% (Fig.~\ref{fig1}). (\romannumeral3) Below $T_s^{\parallel}$ and $T_s^{\perp}$, the susceptibility remains constant with $|\Delta\chi/\chi|$ $<$ 2\% (Fig.~\ref{fig1}), while in a conventional spin glass the ZFC susceptibility should significantly decrease below the freezing point. We note in passing that even in the ac-data the decrease is relatively small, less than 8\% upon going from $\sim$ 0.1 K down to 50 mK~\cite{PhysRevLett.120.087201}. (\romannumeral4) Phenomenologically, there is no $T_c$ $\sim$ 0.1 K scaling behavior of the magnetization upon approaching the anomaly from above~\cite{supple,PhysRevLett.56.416}. Both the magnetic susceptibility and heat capacity show power-law behavior~\cite{vojta2003quantum,PhysRevLett.104.147201,deguchi2012quantum}, $\chi$ $\sim$ ($T$-$T_c$)$^{-\gamma}$ and $C_m$ $\sim$ ($T$-$T_c$)$^{\alpha}$, where $\gamma$ $\sim$ 1/3~\cite{supple}, $\alpha$ $\sim$ 2/3~\cite{li2015gapless}, and $T_c$ = 0 K. (\romannumeral5) The anomaly in the dc-data shifts to higher temperatures upon increasing the magnetic field and follows the Zeeman energies in the zero-field limit~\cite{supple}, similar to paramagnets~\cite{li2018gapped}. In contrast, magnetic field will generally suppress the transition in a conventional spin glass, so that the opposite trend would be observed~\cite{PhysRevLett.110.137201,PhysRevB.95.180411}.

Should the low-$T$ state of YbMgGaO$_4$ be a spin glass, it must be highly unconventional. However, it seems more plausible that a small amount of frozen spin degrees of freedom [$S_m$(0.1 K) $\leq$ 3\%]~\cite{li2015gapless} coexists with the majority remaining dynamic. Moreover, finite zero-temperature susceptibility reveals the presence of gapless low-energy excitations that may be central to the physics of YbMgGaO$_4$. We elucidate their nature below.

\begin{figure*}[t]
\begin{center}
\includegraphics[width=18cm,angle=0]{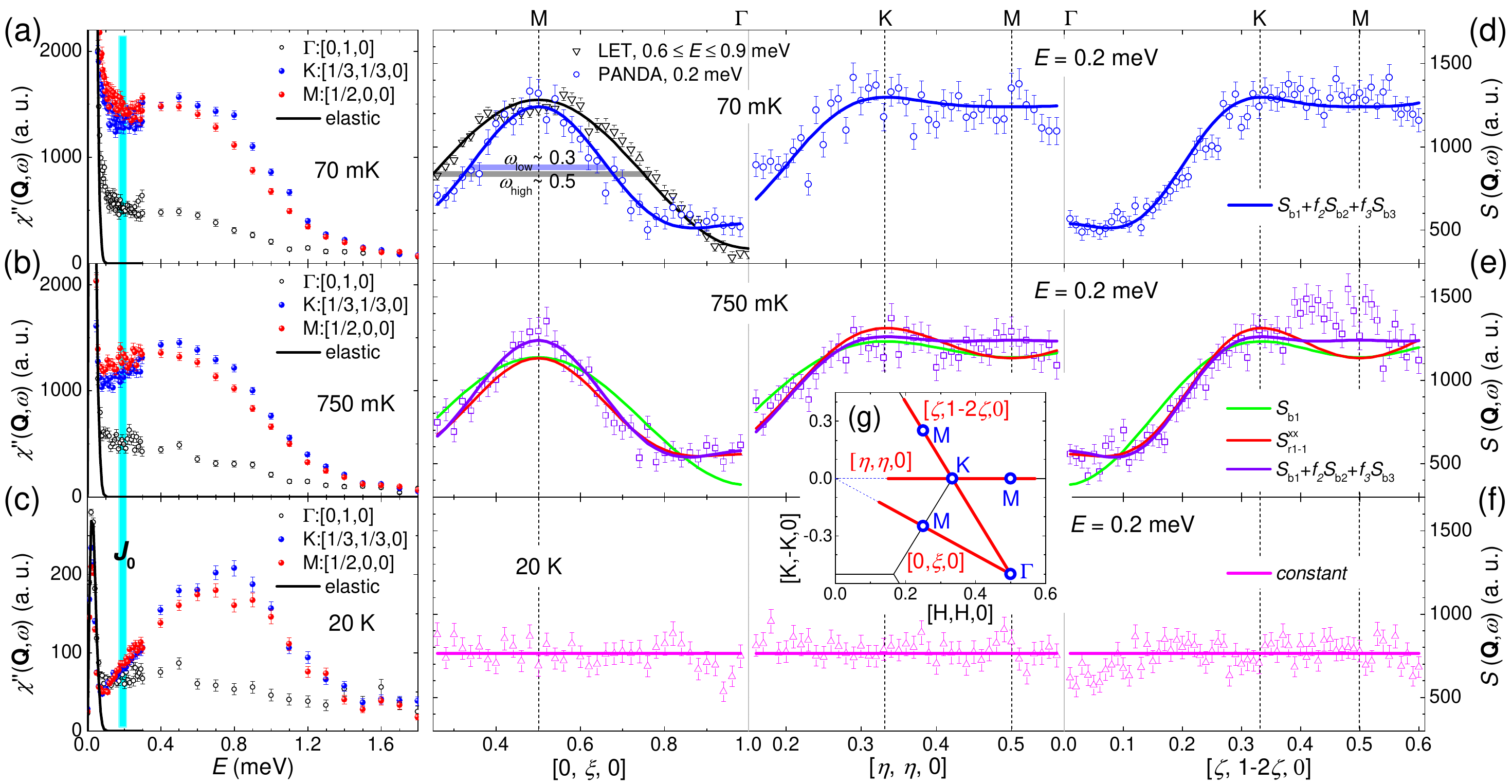}
\caption{(Color online)
Triple-axis INS data of YbMgGaO$_4$ measured on PANDA. Transfer-energy dependence of the dynamical spin susceptibility~\cite{supple} measured at (a) 70 mK, (b) 750 mK, and (c) 20 K, at different high-symmetry \textbf{Q} points. The black lines show the Gaussian fits to the elastic signal at the $\Gamma$ point at -0.05 $\leq E\leq$ 0.05 meV. Wave-vector dependence of the dynamical spin-correlation function~\cite{supple} measured at (d) 70 mK, (e) 750 mK, and (f) 20 K, at the transfer-energy of 0.2 meV. The colored lines present the combined fits to the PANDA data using different model functions (see main text). High-energy (0.6 $\leq$ $E$ $\leq$ 0.9 meV) LET data~\cite{supple} are shown along [0,$\xi$,0] in (d) for comparison, both the low and high energy full widths at half maximum (FWHM) are marked. (g) \textbf{Q}-scans. The red lines show the high-symmetry directions with special \textbf{Q} points labeled, and the black lines represent Brillouin zone boundaries.}
\label{fig2}
\end{center}
\end{figure*}

\emph{High-energy spin excitations.}---Before turning to the low-energy part of the spectrum, let's briefly discuss its high-energy part. Previous time-of-flight LET data suggested that excitations above 0.5 meV can be naturally ascribed to the breaking of NN VBs~\cite{li2017nearest}. Such a high-energy excitation continuum centered at 3$-$6$J_0$ is also clearly observed in the triple-axis measurement on PANDA up to 20 K [see Fig.~\ref{fig2}(a)$-$~\ref{fig2}(c)]. The wave-vector dependence of the spectral weight is well described by the $S_{b1}$ model that includes the breaking of uncorrelated NN VBs, for both the LET data integrated over 0.5 $\leq E\leq$ 1.5 meV~\cite{li2017nearest} and the PANDA data collected at $E$ = 0.7 meV~\cite{supple}. Because a finite energy of $J_0$ is needed to break a singlet, the high-energy continuum clearly shows a gap at $E\leq J_0$ [see Fig.~\ref{fig2}(c)], confirming the scenario of NN VB excitations~\cite{li2017nearest}. The temperature dependence of the dynamical spin susceptibilities roughly follows that of the bulk susceptibilities~\cite{supple}, suggesting that our INS signal is predominantly magnetic.

Recently, the signal observed at high temperatures (10 K) was attributed to the anisotropy of the magnetic form-factor (MFF) of Yb$^{3+}$~\cite{toth2017strong}. As MFF is independent of the transfer energy $E$~\cite{PhysRevB.79.140405}, and high temperature smears the low-energy excitations out, we expect that any \textbf{Q}-dependence observed at high temperatures and low energies should be due to the MFF. However, we find that the INS signal measured at 20 K is almost \textbf{Q}-independent, and shows no anisotropy at $E\leq J_0$ [see Fig.~\ref{fig2}(c) and~\ref{fig2}(f)]. Therefore, the \textbf{Q}-dependent INS signal observed at $E>J_0$ and 20 or 35 K [see Fig.~\ref{fig2}(c)] predominantly originates from the spin-spin correlations, and not from the anisotropy of MFF. This anisotropy is negligible, and the dipole approximation for the MFF is good enough for YbMgGaO$_4$ at $|\mathbf{Q}|\leq$ $4\pi/(\sqrt{3}a)$ $\sim$ 2.2 {\AA}$^{-1}$~\cite{PhysRevB.79.140405}.

\emph{Low-energy spin excitations.}---The temperature-independent spin susceptibility below $T_s^{\parallel}$ and $T_s^{\perp}$ (Fig.~\ref{fig1}) and the power-law behavior of the magnetic heat capacity~\cite{li2015gapless} suggest the presence of distinct gapless low-energy excitations in YbMgGaO$_4$. Unfortunately, these excitations were not well resolved in our previous time-of-flight data~\cite{li2017nearest}, where contamination by the background signal from the magnet posed a serious problem~\cite{supple} that we were able to remedy in the triple-axis experiment performed without the magnet.

\begin{figure*}[t]
\begin{center}
\includegraphics[width=18cm,angle=0]{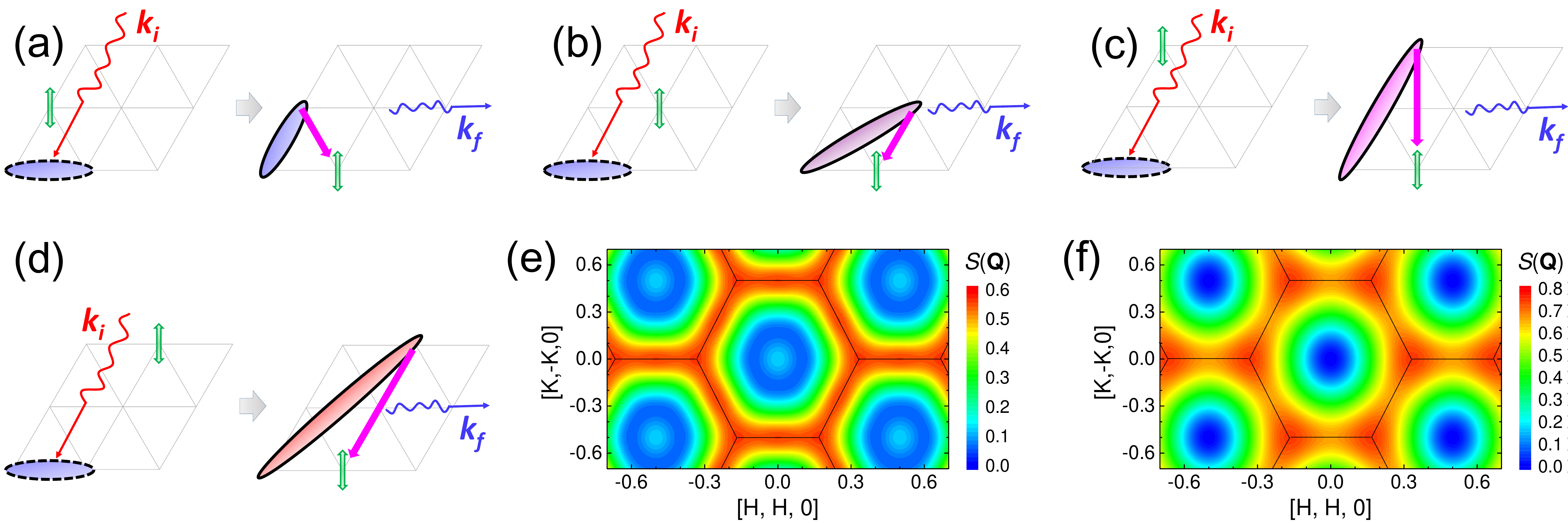}
\caption{(Color online)
Sketch of the neutron scattering upon the rearrangement of NN uncorrelated VBs on the triangular lattice. The nearest-neighbor VB is transformed into another VB between (a) nearest neighbors; (b) second neighbors; (c) third neighbors; (d) fourth neighbors. The ellipses, double arrows, and the magenta arrows present spin singlets, unpaired spins, and the propagation paths of the unpaired spin, respectively. (e) Calculated low-energy correlation function with $f_2$ = -0.22 and $f_3$ = -0.29 (median values from INS measurements). The high-energy counterpart with $f_2$ = $f_3$ = 0 is shown in (f) for comparison. The black lines represent Brillouin zone boundaries.}
\label{fig3}
\end{center}
\end{figure*}

The high-energy excitations are fully gapped, and the INS signal becomes wave-vector independent at 20 K below $\sim$ $J_0$ [see Fig.~\ref{fig2}(c) and~\ref{fig2}(f)]. Conversely, the INS signal at low temperatures (70 and 750 mK) shows a pronounced \textbf{Q}-dependence at $E\leq$ $J_0$ [see Fig.~\ref{fig2}(a),~\ref{fig2}(b),~\ref{fig2}(d), and~\ref{fig2}(e)], which is now characteristic of the low-energy part of the excitation continuum. These excitations are very different from their high-energy counterparts [Fig.~\ref{fig2}(d)]. Indeed, fitting with the same $S_{b1}$ model leads to the least-$R_p$ = 1.44 at 70 mK and least-$R_p$ = 1.37 at 750 mK [Fig.~\ref{fig2}(e)]~\cite{supple}. The most conspicuous differences at low energies are the significantly narrowed peaks in the \textbf{Q}-dependence (along [0,$\xi$,0] for example) and the higher intensity around the $\Gamma$ points [Fig.~\ref{fig2}(d) and~\ref{fig2}(e)]. We notice that the INS signal of the kagome-lattice QSL candidate, ZnCu$_3$(OD)$_6$Cl$_2$, shows very similar features in the low-energy part of the spectrum (i.e., at energies below $J$)~\cite{han2012fractionalized}.

Qualitatively, the narrowing of the \textbf{Q}-dependence indicates an increase in the correlation length of the excitations. For a quantitative analysis we use the equal-time spin correlation function that is normally applied to the spectral weight integrated over energy, but in our case can also be used to fit \textbf{Q}-scans in a narrow energy range, as we deal with excitations that feature very similar \textbf{Q}-dependence at different energies (see~\cite{supple} for further details),
\begin{equation}
S(\mathbf{Q})=A\sum_{j=1}^{\infty}\sum_{l=1}^{Z_j}\frac{f_j}{Z_j}\sin^2\left(\frac{\mathbf{Q}\cdot\mathbf{R}_l^j}{2}\right)+B,
\label{eq1}
\end{equation}
where $A$ is the pre-factor, $B$ is the background due to unpaired spins, $f_j$ is the contribution of the $j^{th}$-neighbors, $Z_j$ is the coordination number for the $j^{th}$-neighbors, and $\mathbf{R}_l^j$ is the position of the $l^{th}$ spin among the $j^{th}$-neighbors.

The coefficients $f_j$ are taken as simple fitting parameters, but on the microscopic level they are related to several processes that involve VBs and unpaired spins~\cite{supple}. In a RVB/QSL state, the low-energy excitations arise from two groups of processes: (\romannumeral1) The neutron breaks a weak VB beyond nearest neighbors, thus producing two unpaired spins. (\romannumeral2) The neutron re-arranges an uncorrelated VB and/or propagates an unpaired spin (see Fig.~\ref{fig3} for example). In the former case, one expects all $f_j$ ($j\geq$ 2)'s positive and $f_1$ = 0, whereas negative $f_j$ ($j\geq$ 2) values and $f_1$ $\neq$ 0 would signal that the processes of the second type contribute to the scattering.

The breaking of VBs beyond nearest neighbors should produce an INS signal with a small periodicity in the \textbf{Q} space, inconsistent with our experimental observations~\cite{supple}. One example would be the recent phenomenological model by Kimchi \emph{et al.} who proposed $f_1$:$f_2$ = 4:1 and $f_j\equiv$ 0 ($j\geq$ 3) [see Eq.~(\ref{eq1})] for the low-energy excitations~\cite{kimchi2017valence}. This model successfully explains the relative increase in the INS intensity around the M points below $\sim$ $J_0$ [see Fig.~\ref{fig2}(a) and~\ref{fig2}(b)], but does not account for the narrowing of the \textbf{Q}-dependence and for the increased intensity around the $\Gamma$ points~\cite{supple}.

Therefore, the processes of the second type should be taken into account. The constant susceptibility in the zero-temperature limit would suggest the presence of unpaired spins not bound into VBs, so we focus on those processes that combine the re-arrangement of the VBs with the propagation of unpaired spins. They are exemplified in Fig.~\ref{fig3} that shows the $S_{r1-1}^{xx}$ model with $f_2$ = -1/4, $f_3$ = -1/8, and $f_j$ $\equiv$ 0 ($j\geq$ 4, here we fix $f_1$ $\equiv$ 1) [see Eq.~(\ref{eq1})]~\cite{supple}.
This model describes the experimental low-energy INS data much better than the $S_{b1}$ model, with the least-$R_p$ = 1.00 at 70 mK and least-$R_p$ = 1.03 at 750 mK [Fig.~\ref{fig2}(e)], respectively.

The best description of the experimental data is achieved by treating $f_j$ ($j\geq$ 2) as free parameters while fixing $f_1$ $\equiv$ 1. We find that two such parameters, $f_2$ and $f_3$, are sufficient to fit the spectra. The addition of $f_4$ does not improve the fits significantly~\footnote{The $S_{b1}+f_2S_{b2}+f_3S_{b3}+f_4S_{b4}$ fits yield $|f_4|$ $<$ 0.04, and the least-$R_p$ decreases by less than 0.02, compared to the $S_{b1}+f_2S_{b2}+f_3S_{b3}$ fits}, indicating a relatively short correlation length, $\xi_{low}$ $\sim$ 2$a$ = 6.8 {\AA}~\cite{supple}. Through the $S_{b1}+f_2S_{b2}+f_3S_{b3}$ fits, we get $f_2$ = -0.18, $f_3$ = -0.21 with the least-$R_p$ = 0.89 at 70 mK [Fig.~\ref{fig2}(d)] and $f_2$ = -0.18, $f_3$ = -0.24 with the least-$R_p$ = 0.87 at 750 mK [Fig.~\ref{fig2}(e)]. For the quasielastic neutron scattering at 0.07 meV, the very similar results are obtained~\cite{supple}. The fitted values of $f_2$ and $f_3$ are relatively close to those expected for the scattering process depicted in Fig.~\ref{fig3}(a), suggesting that the re-arrangement of VBs and the propagation of unpaired spins make a significant contribution to the low-energy excitations.

\emph{Discussion.}---In Ref.~\cite{li2017nearest}, we conjectured that the high-energy part of the excitation continuum arising from the breaking of NN VBs should be preceded by the distinct low-energy part driven by excitations of different nature. We are now able to confirm directly that the \textbf{Q}-dependence of the spectral weight changes indeed. At low energies, spin-spin correlations extend beyond nearest neighbors and can be represented by the re-arrangement of VBs that facilitates the propagation of unpaired spins. This is reminiscent of the original concept of fermionic excitations~\cite{shen2016spinon}, although the propagation of unpaired spins in YbMgGaO$_4$ must be limited, as no magnetic contribution to the thermal conductivity has been observed~\cite{PhysRevLett.117.267202}. We suggest that random magnetic couplings caused by the mixing of Mg$^{2+}$ and Ga$^{3+}$ may restrict the propagation of unpaired spins. We also note that unpaired spins are integral to the ground state of YbMgGaO$_4$, and their concentration estimated from the size of the INS background~\cite{supple} is significantly larger than the fraction of spins that may become frozen around 0.1 K.

On a related note, we mention that a similar description of low-energy excitations holds for the QSL candidate herbertsmithite (ZnCu$_3$(OD)$_6$Cl$_2$) that features spins on the kagome lattice~\cite{han2012fractionalized,supple}. Moreover, the similar narrowing of the excitation continuum at low energies was also reported in the one-dimensional KCuF$_3$~\cite{lake2005quantum,PhysRevLett.111.137205} and honeycomb Kitaev system $\alpha$-RuCl$_3$~\cite{banerjee2017neutron,do2017majorana}.

\emph{Conclusions.}---The low-$T$ magnetization data and low-energy neutron spectroscopy help us to resolve several open issues in the putative spin-liquid physics of YbMgGaO$_4$. Our observations suggest that only an insignificant fraction of spin degrees of freedom may become static, whereas dynamics of the majority gives rise to the broad excitation continuum. We interpret this continuum as consisting of two parts, the breaking of nearest-neighbor VBs at high energies and the re-arrangement of VBs at low energies. Although the low-energy excitations we observed are rather similar to the anticipated fermionic excitations, the propagation of unpaired spins must be curtailed. More generally, we argue that a similar formalism of the low-energy rearrangement of VBs is applicable to herbertsmithite and may be a universal feature of spin-liquid states with unusual spin dynamics, a problem that clearly warrants further theoretical investigation.

\acknowledgments
\emph{Acknowledgment.}---We thank Yixi Su, Erxi Feng, Hao Deng, Fengfeng Zhu, and Junda Song for helpful discussions. B. L. was supported by China Scholarship Council, the National Natural Science Foundation of China (No. 11875238), and Science Challenge Project (No. TZ2016004). The work in Augsburg was supported by the German Science Foundation through TRR-80 and by the German Federal Ministry for Education and Research through the Sofja Kovalevskaya Award of the Alexander von Humboldt Foundation.

\bibliography{yb}

\begin{thebibliography}{46}%
\makeatletter
\providecommand \@ifxundefined [1]{%
 \@ifx{#1\undefined}
}%
\providecommand \@ifnum [1]{%
 \ifnum #1\expandafter \@firstoftwo
 \else \expandafter \@secondoftwo
 \fi
}%
\providecommand \@ifx [1]{%
 \ifx #1\expandafter \@firstoftwo
 \else \expandafter \@secondoftwo
 \fi
}%
\providecommand \natexlab [1]{#1}%
\providecommand \enquote  [1]{``#1''}%
\providecommand \bibnamefont  [1]{#1}%
\providecommand \bibfnamefont [1]{#1}%
\providecommand \citenamefont [1]{#1}%
\providecommand \href@noop [0]{\@secondoftwo}%
\providecommand \href [0]{\begingroup \@sanitize@url \@href}%
\providecommand \@href[1]{\@@startlink{#1}\@@href}%
\providecommand \@@href[1]{\endgroup#1\@@endlink}%
\providecommand \@sanitize@url [0]{\catcode `\\12\catcode `\$12\catcode
  `\&12\catcode `\#12\catcode `\^12\catcode `\_12\catcode `\%12\relax}%
\providecommand \@@startlink[1]{}%
\providecommand \@@endlink[0]{}%
\providecommand \url  [0]{\begingroup\@sanitize@url \@url }%
\providecommand \@url [1]{\endgroup\@href {#1}{\urlprefix }}%
\providecommand \urlprefix  [0]{URL }%
\providecommand \Eprint [0]{\href }%
\providecommand \doibase [0]{http://dx.doi.org/}%
\providecommand \selectlanguage [0]{\@gobble}%
\providecommand \bibinfo  [0]{\@secondoftwo}%
\providecommand \bibfield  [0]{\@secondoftwo}%
\providecommand \translation [1]{[#1]}%
\providecommand \BibitemOpen [0]{}%
\providecommand \bibitemStop [0]{}%
\providecommand \bibitemNoStop [0]{.\EOS\space}%
\providecommand \EOS [0]{\spacefactor3000\relax}%
\providecommand \BibitemShut  [1]{\csname bibitem#1\endcsname}%
\let\auto@bib@innerbib\@empty
\bibitem [{\citenamefont {Anderson}(1973)}]{anderson1973resonating}%
  \BibitemOpen
  \bibfield  {author} {\bibinfo {author} {\bibfnamefont {P.~W.}\ \bibnamefont
  {Anderson}},\ }\bibfield  {title} {\enquote {\bibinfo {title} {Resonating
  valence bonds: {A} new kind of insulator?}}\ }\href@noop {} {\bibfield
  {journal} {\bibinfo  {journal} {Mater. Res. Bull.}\ }\textbf {\bibinfo
  {volume} {8}},\ \bibinfo {pages} {153--160} (\bibinfo {year}
  {1973})}\BibitemShut {NoStop}%
\bibitem [{\citenamefont {Anderson}(1987)}]{anderson1987resonating}%
  \BibitemOpen
  \bibfield  {author} {\bibinfo {author} {\bibfnamefont {P.~W.}\ \bibnamefont
  {Anderson}},\ }\bibfield  {title} {\enquote {\bibinfo {title} {The resonating
  valence bond state in {La$_2$CuO$_4$} and superconductivity},}\ }\href@noop
  {} {\bibfield  {journal} {\bibinfo  {journal} {Science}\ }\textbf {\bibinfo
  {volume} {235}},\ \bibinfo {pages} {1196--1198} (\bibinfo {year}
  {1987})}\BibitemShut {NoStop}%
\bibitem [{\citenamefont {Nayak}\ \emph {et~al.}(2008)\citenamefont {Nayak},
  \citenamefont {Simon}, \citenamefont {Stern}, \citenamefont {Freedman},\ and\
  \citenamefont {Sarma}}]{nayak2008non}%
  \BibitemOpen
  \bibfield  {author} {\bibinfo {author} {\bibfnamefont {C.}~\bibnamefont
  {Nayak}}, \bibinfo {author} {\bibfnamefont {S.~H.}\ \bibnamefont {Simon}},
  \bibinfo {author} {\bibfnamefont {A.}~\bibnamefont {Stern}}, \bibinfo
  {author} {\bibfnamefont {M.}~\bibnamefont {Freedman}}, \ and\ \bibinfo
  {author} {\bibfnamefont {S.~D.}\ \bibnamefont {Sarma}},\ }\bibfield  {title}
  {\enquote {\bibinfo {title} {Non-abelian anyons and topological quantum
  computation},}\ }\href@noop {} {\bibfield  {journal} {\bibinfo  {journal}
  {Rev. Mod. Phys.}\ }\textbf {\bibinfo {volume} {80}},\ \bibinfo {pages}
  {1083} (\bibinfo {year} {2008})}\BibitemShut {NoStop}%
\bibitem [{\citenamefont {Wen}(2004)}]{wen2004quantum}%
  \BibitemOpen
  \bibfield  {author} {\bibinfo {author} {\bibfnamefont {X.-G.}\ \bibnamefont
  {Wen}},\ }\href@noop {} {\emph {\bibinfo {title} {Quantum field theory of
  many-body systems: from the origin of sound to an origin of light and
  electrons}}}\ (\bibinfo  {publisher} {Oxford University Press on Demand},\
  \bibinfo {year} {2004})\BibitemShut {NoStop}%
\bibitem [{\citenamefont {Moessner}\ and\ \citenamefont
  {Ramirez}(2006)}]{moessner2006geometrical}%
  \BibitemOpen
  \bibfield  {author} {\bibinfo {author} {\bibfnamefont {R.}~\bibnamefont
  {Moessner}}\ and\ \bibinfo {author} {\bibfnamefont {A.~P.}\ \bibnamefont
  {Ramirez}},\ }\bibfield  {title} {\enquote {\bibinfo {title} {Geometrical
  frustration},}\ }\href@noop {} {\bibfield  {journal} {\bibinfo  {journal}
  {Phys. Today}\ }\textbf {\bibinfo {volume} {59}},\ \bibinfo {pages} {24}
  (\bibinfo {year} {2006})}\BibitemShut {NoStop}%
\bibitem [{\citenamefont {Lee}(2008)}]{lee2008end}%
  \BibitemOpen
  \bibfield  {author} {\bibinfo {author} {\bibfnamefont {P.~A.}\ \bibnamefont
  {Lee}},\ }\bibfield  {title} {\enquote {\bibinfo {title} {An end to the
  drought of quantum spin liquids},}\ }\href@noop {} {\bibfield  {journal}
  {\bibinfo  {journal} {Science}\ }\textbf {\bibinfo {volume} {321}},\ \bibinfo
  {pages} {1306--1307} (\bibinfo {year} {2008})}\BibitemShut {NoStop}%
\bibitem [{\citenamefont {Balents}(2010)}]{balents2010spin}%
  \BibitemOpen
  \bibfield  {author} {\bibinfo {author} {\bibfnamefont {L.}~\bibnamefont
  {Balents}},\ }\bibfield  {title} {\enquote {\bibinfo {title} {Spin liquids in
  frustrated magnets},}\ }\href@noop {} {\bibfield  {journal} {\bibinfo
  {journal} {Nature}\ }\textbf {\bibinfo {volume} {464}},\ \bibinfo {pages}
  {199--208} (\bibinfo {year} {2010})}\BibitemShut {NoStop}%
\bibitem [{\citenamefont {Li}\ \emph {et~al.}(2015{\natexlab{a}})\citenamefont
  {Li}, \citenamefont {Liao}, \citenamefont {Zhang}, \citenamefont {Li},
  \citenamefont {Jin}, \citenamefont {Ling}, \citenamefont {Zhang},
  \citenamefont {Zou}, \citenamefont {Pi}, \citenamefont {Yang}, \citenamefont
  {Wang}, \citenamefont {Wu},\ and\ \citenamefont {Zhang}}]{li2015gapless}%
  \BibitemOpen
  \bibfield  {author} {\bibinfo {author} {\bibfnamefont {Y.}~\bibnamefont
  {Li}}, \bibinfo {author} {\bibfnamefont {H.}~\bibnamefont {Liao}}, \bibinfo
  {author} {\bibfnamefont {Z.}~\bibnamefont {Zhang}}, \bibinfo {author}
  {\bibfnamefont {S.}~\bibnamefont {Li}}, \bibinfo {author} {\bibfnamefont
  {F.}~\bibnamefont {Jin}}, \bibinfo {author} {\bibfnamefont {L.}~\bibnamefont
  {Ling}}, \bibinfo {author} {\bibfnamefont {L.}~\bibnamefont {Zhang}},
  \bibinfo {author} {\bibfnamefont {Y.}~\bibnamefont {Zou}}, \bibinfo {author}
  {\bibfnamefont {L.}~\bibnamefont {Pi}}, \bibinfo {author} {\bibfnamefont
  {Z.}~\bibnamefont {Yang}}, \bibinfo {author} {\bibfnamefont {J.}~\bibnamefont
  {Wang}}, \bibinfo {author} {\bibfnamefont {Z.}~\bibnamefont {Wu}}, \ and\
  \bibinfo {author} {\bibfnamefont {Q.}~\bibnamefont {Zhang}},\ }\bibfield
  {title} {\enquote {\bibinfo {title} {Gapless quantum spin liquid ground state
  in the two-dimensional spin-1/2 triangular antiferromagnet {YbMgGaO$_4$}},}\
  }\href@noop {} {\bibfield  {journal} {\bibinfo  {journal} {Sci. Rep.}\
  }\textbf {\bibinfo {volume} {5}},\ \bibinfo {pages} {16419} (\bibinfo {year}
  {2015}{\natexlab{a}})}\BibitemShut {NoStop}%
\bibitem [{Note1()}]{Note1}%
  \BibitemOpen
  \bibinfo {note} {Refs.~\cite {li2015gapless},~\cite {li2015rare}, and~\cite
  {paddison2016continuous} report $J_0$ $\sim $ 0.24, 0.13, and 0.20 meV,
  respectively. Therefore, we use the median value of $J_0$ $\sim $ 0.2 meV
  throughout the Letter.}\BibitemShut {Stop}%
\bibitem [{\citenamefont {Li}\ \emph {et~al.}(2015{\natexlab{b}})\citenamefont
  {Li}, \citenamefont {Chen}, \citenamefont {Tong}, \citenamefont {Pi},
  \citenamefont {Liu}, \citenamefont {Yang}, \citenamefont {Wang},\ and\
  \citenamefont {Zhang}}]{li2015rare}%
  \BibitemOpen
  \bibfield  {author} {\bibinfo {author} {\bibfnamefont {Y.}~\bibnamefont
  {Li}}, \bibinfo {author} {\bibfnamefont {G.}~\bibnamefont {Chen}}, \bibinfo
  {author} {\bibfnamefont {W.}~\bibnamefont {Tong}}, \bibinfo {author}
  {\bibfnamefont {L.}~\bibnamefont {Pi}}, \bibinfo {author} {\bibfnamefont
  {J.}~\bibnamefont {Liu}}, \bibinfo {author} {\bibfnamefont {Z.}~\bibnamefont
  {Yang}}, \bibinfo {author} {\bibfnamefont {X.}~\bibnamefont {Wang}}, \ and\
  \bibinfo {author} {\bibfnamefont {Q.}~\bibnamefont {Zhang}},\ }\bibfield
  {title} {\enquote {\bibinfo {title} {Rare-earth triangular lattice spin
  liquid: a single-crystal study of {YbMgGaO$_4$}},}\ }\href@noop {} {\bibfield
   {journal} {\bibinfo  {journal} {Phys. Rev. Lett.}\ }\textbf {\bibinfo
  {volume} {115}},\ \bibinfo {pages} {167203} (\bibinfo {year}
  {2015}{\natexlab{b}})}\BibitemShut {NoStop}%
\bibitem [{\citenamefont {Xu}\ \emph {et~al.}(2016)\citenamefont {Xu},
  \citenamefont {Zhang}, \citenamefont {Li}, \citenamefont {Yu}, \citenamefont
  {Hong}, \citenamefont {Zhang},\ and\ \citenamefont
  {Li}}]{PhysRevLett.117.267202}%
  \BibitemOpen
  \bibfield  {author} {\bibinfo {author} {\bibfnamefont {Y.}~\bibnamefont
  {Xu}}, \bibinfo {author} {\bibfnamefont {J.}~\bibnamefont {Zhang}}, \bibinfo
  {author} {\bibfnamefont {Y.~S.}\ \bibnamefont {Li}}, \bibinfo {author}
  {\bibfnamefont {Y.~J.}\ \bibnamefont {Yu}}, \bibinfo {author} {\bibfnamefont
  {X.~C.}\ \bibnamefont {Hong}}, \bibinfo {author} {\bibfnamefont {Q.~M.}\
  \bibnamefont {Zhang}}, \ and\ \bibinfo {author} {\bibfnamefont {S.~Y.}\
  \bibnamefont {Li}},\ }\bibfield  {title} {\enquote {\bibinfo {title} {Absence
  of magnetic thermal conductivity in the quantum spin-liquid candidate
  {YbMgGaO$_4$}},}\ }\href@noop {} {\bibfield  {journal} {\bibinfo  {journal}
  {Phys. Rev. Lett.}\ }\textbf {\bibinfo {volume} {117}},\ \bibinfo {pages}
  {267202} (\bibinfo {year} {2016})}\BibitemShut {NoStop}%
\bibitem [{\citenamefont {Li}\ \emph {et~al.}(2016)\citenamefont {Li},
  \citenamefont {Adroja}, \citenamefont {Biswas}, \citenamefont {Baker},
  \citenamefont {Zhang}, \citenamefont {Liu}, \citenamefont {Tsirlin},
  \citenamefont {Gegenwart},\ and\ \citenamefont
  {Zhang}}]{PhysRevLett.117.097201}%
  \BibitemOpen
  \bibfield  {author} {\bibinfo {author} {\bibfnamefont {Y.}~\bibnamefont
  {Li}}, \bibinfo {author} {\bibfnamefont {D.}~\bibnamefont {Adroja}}, \bibinfo
  {author} {\bibfnamefont {P.~K.}\ \bibnamefont {Biswas}}, \bibinfo {author}
  {\bibfnamefont {P.~J.}\ \bibnamefont {Baker}}, \bibinfo {author}
  {\bibfnamefont {Q.}~\bibnamefont {Zhang}}, \bibinfo {author} {\bibfnamefont
  {J.}~\bibnamefont {Liu}}, \bibinfo {author} {\bibfnamefont {A.~A.}\
  \bibnamefont {Tsirlin}}, \bibinfo {author} {\bibfnamefont {P.}~\bibnamefont
  {Gegenwart}}, \ and\ \bibinfo {author} {\bibfnamefont {Q.}~\bibnamefont
  {Zhang}},\ }\bibfield  {title} {\enquote {\bibinfo {title} {Muon spin
  relaxation evidence for the {U(1)} quantum spin-liquid ground state in the
  triangular antiferromagnet {YbMgGaO$_{4}$}},}\ }\href@noop {} {\bibfield
  {journal} {\bibinfo  {journal} {Phys. Rev. Lett.}\ }\textbf {\bibinfo
  {volume} {117}},\ \bibinfo {pages} {097201} (\bibinfo {year}
  {2016})}\BibitemShut {NoStop}%
\bibitem [{\citenamefont {Shen}\ \emph {et~al.}(2016)\citenamefont {Shen},
  \citenamefont {Li}, \citenamefont {Wo}, \citenamefont {Li}, \citenamefont
  {Shen}, \citenamefont {Pan}, \citenamefont {Wang}, \citenamefont {Walker},
  \citenamefont {Steffens}, \citenamefont {Boehm}, \citenamefont {Hao},
  \citenamefont {Quintero-Castro}, \citenamefont {Harriger}, \citenamefont
  {Frontzek}, \citenamefont {Hao}, \citenamefont {Meng}, \citenamefont {Zhang},
  \citenamefont {Chen},\ and\ \citenamefont {Zhao}}]{shen2016spinon}%
  \BibitemOpen
  \bibfield  {author} {\bibinfo {author} {\bibfnamefont {Y.}~\bibnamefont
  {Shen}}, \bibinfo {author} {\bibfnamefont {Y.}~\bibnamefont {Li}}, \bibinfo
  {author} {\bibfnamefont {H.}~\bibnamefont {Wo}}, \bibinfo {author}
  {\bibfnamefont {Y.}~\bibnamefont {Li}}, \bibinfo {author} {\bibfnamefont
  {S.}~\bibnamefont {Shen}}, \bibinfo {author} {\bibfnamefont {B.}~\bibnamefont
  {Pan}}, \bibinfo {author} {\bibfnamefont {Q.}~\bibnamefont {Wang}}, \bibinfo
  {author} {\bibfnamefont {H.~C.}\ \bibnamefont {Walker}}, \bibinfo {author}
  {\bibfnamefont {P.}~\bibnamefont {Steffens}}, \bibinfo {author}
  {\bibfnamefont {M.}~\bibnamefont {Boehm}}, \bibinfo {author} {\bibfnamefont
  {Y.}~\bibnamefont {Hao}}, \bibinfo {author} {\bibfnamefont {D.~L.}\
  \bibnamefont {Quintero-Castro}}, \bibinfo {author} {\bibfnamefont {L.~W.}\
  \bibnamefont {Harriger}}, \bibinfo {author} {\bibfnamefont {M.~D.}\
  \bibnamefont {Frontzek}}, \bibinfo {author} {\bibfnamefont {L.}~\bibnamefont
  {Hao}}, \bibinfo {author} {\bibfnamefont {S.}~\bibnamefont {Meng}}, \bibinfo
  {author} {\bibfnamefont {Q.}~\bibnamefont {Zhang}}, \bibinfo {author}
  {\bibfnamefont {G}~\bibnamefont {Chen}}, \ and\ \bibinfo {author}
  {\bibfnamefont {J.}~\bibnamefont {Zhao}},\ }\bibfield  {title} {\enquote
  {\bibinfo {title} {Evidence for a spinon {Fermi} surface in a
  triangular-lattice quantum-spin-liquid candidate},}\ }\href@noop {}
  {\bibfield  {journal} {\bibinfo  {journal} {Nature}\ }\textbf {\bibinfo
  {volume} {540}},\ \bibinfo {pages} {559--562} (\bibinfo {year}
  {2016})}\BibitemShut {NoStop}%
\bibitem [{\citenamefont {Li}\ \emph {et~al.}(2017{\natexlab{a}})\citenamefont
  {Li}, \citenamefont {Adroja}, \citenamefont {Bewley}, \citenamefont
  {Voneshen}, \citenamefont {Tsirlin}, \citenamefont {Gegenwart},\ and\
  \citenamefont {Zhang}}]{PhysRevLett.118.107202}%
  \BibitemOpen
  \bibfield  {author} {\bibinfo {author} {\bibfnamefont {Y.}~\bibnamefont
  {Li}}, \bibinfo {author} {\bibfnamefont {D.}~\bibnamefont {Adroja}}, \bibinfo
  {author} {\bibfnamefont {R.~I.}\ \bibnamefont {Bewley}}, \bibinfo {author}
  {\bibfnamefont {D.}~\bibnamefont {Voneshen}}, \bibinfo {author}
  {\bibfnamefont {A.~A.}\ \bibnamefont {Tsirlin}}, \bibinfo {author}
  {\bibfnamefont {P.}~\bibnamefont {Gegenwart}}, \ and\ \bibinfo {author}
  {\bibfnamefont {Q.}~\bibnamefont {Zhang}},\ }\bibfield  {title} {\enquote
  {\bibinfo {title} {Crystalline electric-field randomness in the triangular
  lattice spin-liquid {YbMgGaO$_{4}$}},}\ }\href@noop {} {\bibfield  {journal}
  {\bibinfo  {journal} {Phys. Rev. Lett.}\ }\textbf {\bibinfo {volume} {118}},\
  \bibinfo {pages} {107202} (\bibinfo {year} {2017}{\natexlab{a}})}\BibitemShut
  {NoStop}%
\bibitem [{\citenamefont {Paddison}\ \emph {et~al.}(2017)\citenamefont
  {Paddison}, \citenamefont {Daum}, \citenamefont {Dun}, \citenamefont
  {Ehlers}, \citenamefont {Liu}, \citenamefont {Stone}, \citenamefont {Zhou},\
  and\ \citenamefont {Mourigal}}]{paddison2016continuous}%
  \BibitemOpen
  \bibfield  {author} {\bibinfo {author} {\bibfnamefont {J.~A.~M.}\
  \bibnamefont {Paddison}}, \bibinfo {author} {\bibfnamefont {M.}~\bibnamefont
  {Daum}}, \bibinfo {author} {\bibfnamefont {Z.}~\bibnamefont {Dun}}, \bibinfo
  {author} {\bibfnamefont {G.}~\bibnamefont {Ehlers}}, \bibinfo {author}
  {\bibfnamefont {Y.}~\bibnamefont {Liu}}, \bibinfo {author} {\bibfnamefont
  {M.~B.}\ \bibnamefont {Stone}}, \bibinfo {author} {\bibfnamefont
  {H.}~\bibnamefont {Zhou}}, \ and\ \bibinfo {author} {\bibfnamefont
  {M.}~\bibnamefont {Mourigal}},\ }\bibfield  {title} {\enquote {\bibinfo
  {title} {Continuous excitations of the triangular-lattice quantum spin liquid
  {YbMgGaO$_4$}},}\ }\href@noop {} {\bibfield  {journal} {\bibinfo  {journal}
  {Nat. Phys.}\ }\textbf {\bibinfo {volume} {13}},\ \bibinfo {pages} {117--122}
  (\bibinfo {year} {2017})}\BibitemShut {NoStop}%
\bibitem [{\citenamefont {Zhang}\ \emph {et~al.}(2018)\citenamefont {Zhang},
  \citenamefont {Mahmood}, \citenamefont {Daum}, \citenamefont {Dun},
  \citenamefont {Paddison}, \citenamefont {Laurita}, \citenamefont {Hong},
  \citenamefont {Zhou}, \citenamefont {Armitage},\ and\ \citenamefont
  {Mourigal}}]{zhang2017hierarchy}%
  \BibitemOpen
  \bibfield  {author} {\bibinfo {author} {\bibfnamefont {X.}~\bibnamefont
  {Zhang}}, \bibinfo {author} {\bibfnamefont {F.}~\bibnamefont {Mahmood}},
  \bibinfo {author} {\bibfnamefont {M.}~\bibnamefont {Daum}}, \bibinfo {author}
  {\bibfnamefont {Z.}~\bibnamefont {Dun}}, \bibinfo {author} {\bibfnamefont
  {J.~A.~M.}\ \bibnamefont {Paddison}}, \bibinfo {author} {\bibfnamefont
  {N.~J.}\ \bibnamefont {Laurita}}, \bibinfo {author} {\bibfnamefont
  {T.}~\bibnamefont {Hong}}, \bibinfo {author} {\bibfnamefont {H.}~\bibnamefont
  {Zhou}}, \bibinfo {author} {\bibfnamefont {N.~P.}\ \bibnamefont {Armitage}},
  \ and\ \bibinfo {author} {\bibfnamefont {M.}~\bibnamefont {Mourigal}},\
  }\bibfield  {title} {\enquote {\bibinfo {title} {Hierarchy of exchange
  interactions in the triangular-lattice spin-liquid {YbMgGaO$_4$}},}\ }\href
  {\doibase 10.1103/PhysRevX.8.031001} {\bibfield  {journal} {\bibinfo
  {journal} {Phys. Rev. X}\ }\textbf {\bibinfo {volume} {8}},\ \bibinfo {pages}
  {031001} (\bibinfo {year} {2018})}\BibitemShut {NoStop}%
\bibitem [{\citenamefont {Ma}\ \emph {et~al.}(2018)\citenamefont {Ma},
  \citenamefont {Wang}, \citenamefont {Dong}, \citenamefont {Zhang},
  \citenamefont {Li}, \citenamefont {Zheng}, \citenamefont {Yu}, \citenamefont
  {Wang}, \citenamefont {Che}, \citenamefont {Ran}, \citenamefont {Bao},
  \citenamefont {Cai}, \citenamefont {\ifmmode~\check{C}\else
  \v{C}\fi{}erm\'ak}, \citenamefont {Schneidewind}, \citenamefont {Yano},
  \citenamefont {Gardner}, \citenamefont {Lu}, \citenamefont {Yu},
  \citenamefont {Liu}, \citenamefont {Li}, \citenamefont {Li},\ and\
  \citenamefont {Wen}}]{PhysRevLett.120.087201}%
  \BibitemOpen
  \bibfield  {author} {\bibinfo {author} {\bibfnamefont {Z.}~\bibnamefont
  {Ma}}, \bibinfo {author} {\bibfnamefont {J.}~\bibnamefont {Wang}}, \bibinfo
  {author} {\bibfnamefont {Z.-Y.}\ \bibnamefont {Dong}}, \bibinfo {author}
  {\bibfnamefont {J.}~\bibnamefont {Zhang}}, \bibinfo {author} {\bibfnamefont
  {S.}~\bibnamefont {Li}}, \bibinfo {author} {\bibfnamefont {S.-H.}\
  \bibnamefont {Zheng}}, \bibinfo {author} {\bibfnamefont {Y.}~\bibnamefont
  {Yu}}, \bibinfo {author} {\bibfnamefont {W.}~\bibnamefont {Wang}}, \bibinfo
  {author} {\bibfnamefont {L.}~\bibnamefont {Che}}, \bibinfo {author}
  {\bibfnamefont {K.}~\bibnamefont {Ran}}, \bibinfo {author} {\bibfnamefont
  {S.}~\bibnamefont {Bao}}, \bibinfo {author} {\bibfnamefont {Z.}~\bibnamefont
  {Cai}}, \bibinfo {author} {\bibfnamefont {P.}~\bibnamefont
  {\ifmmode~\check{C}\else \v{C}\fi{}erm\'ak}}, \bibinfo {author}
  {\bibfnamefont {A.}~\bibnamefont {Schneidewind}}, \bibinfo {author}
  {\bibfnamefont {S.}~\bibnamefont {Yano}}, \bibinfo {author} {\bibfnamefont
  {J.~S.}\ \bibnamefont {Gardner}}, \bibinfo {author} {\bibfnamefont
  {X.}~\bibnamefont {Lu}}, \bibinfo {author} {\bibfnamefont {S.-L.}\
  \bibnamefont {Yu}}, \bibinfo {author} {\bibfnamefont {J.-M.}\ \bibnamefont
  {Liu}}, \bibinfo {author} {\bibfnamefont {S.}~\bibnamefont {Li}}, \bibinfo
  {author} {\bibfnamefont {J.-X.}\ \bibnamefont {Li}}, \ and\ \bibinfo {author}
  {\bibfnamefont {J.}~\bibnamefont {Wen}},\ }\bibfield  {title} {\enquote
  {\bibinfo {title} {Spin-glass ground state in a triangular-lattice compound
  {YbZnGaO$_4$}},}\ }\href@noop {} {\bibfield  {journal} {\bibinfo  {journal}
  {Phys. Rev. Lett.}\ }\textbf {\bibinfo {volume} {120}},\ \bibinfo {pages}
  {087201} (\bibinfo {year} {2018})}\BibitemShut {NoStop}%
\bibitem [{\citenamefont {Zhu}\ \emph {et~al.}(2017)\citenamefont {Zhu},
  \citenamefont {Maksimov}, \citenamefont {White},\ and\ \citenamefont
  {Chernyshev}}]{PhysRevLett.119.157201}%
  \BibitemOpen
  \bibfield  {author} {\bibinfo {author} {\bibfnamefont {Z.}~\bibnamefont
  {Zhu}}, \bibinfo {author} {\bibfnamefont {P.~A.}\ \bibnamefont {Maksimov}},
  \bibinfo {author} {\bibfnamefont {S.~R.}\ \bibnamefont {White}}, \ and\
  \bibinfo {author} {\bibfnamefont {A.~L.}\ \bibnamefont {Chernyshev}},\
  }\bibfield  {title} {\enquote {\bibinfo {title} {Disorder-induced mimicry of
  a spin liquid in {YbMgGaO$_4$}},}\ }\href@noop {} {\bibfield  {journal}
  {\bibinfo  {journal} {Phys. Rev. Lett.}\ }\textbf {\bibinfo {volume} {119}},\
  \bibinfo {pages} {157201} (\bibinfo {year} {2017})}\BibitemShut {NoStop}%
\bibitem [{\citenamefont {Kimchi}\ \emph {et~al.}(2018)\citenamefont {Kimchi},
  \citenamefont {Nahum},\ and\ \citenamefont {Senthil}}]{kimchi2017valence}%
  \BibitemOpen
  \bibfield  {author} {\bibinfo {author} {\bibfnamefont {I.}~\bibnamefont
  {Kimchi}}, \bibinfo {author} {\bibfnamefont {A.}~\bibnamefont {Nahum}}, \
  and\ \bibinfo {author} {\bibfnamefont {T.}~\bibnamefont {Senthil}},\
  }\bibfield  {title} {\enquote {\bibinfo {title} {Valence bonds in random
  quantum magnets: {Theory} and application to {YbMgGaO$_4$}},}\ }\href
  {\doibase 10.1103/PhysRevX.8.031028} {\bibfield  {journal} {\bibinfo
  {journal} {Phys. Rev. X}\ }\textbf {\bibinfo {volume} {8}},\ \bibinfo {pages}
  {031028} (\bibinfo {year} {2018})}\BibitemShut {NoStop}%
\bibitem [{\citenamefont {Li}\ \emph {et~al.}(2017{\natexlab{b}})\citenamefont
  {Li}, \citenamefont {Adroja}, \citenamefont {Voneshen}, \citenamefont
  {Bewley}, \citenamefont {Zhang}, \citenamefont {Tsirlin},\ and\ \citenamefont
  {Gegenwart}}]{li2017nearest}%
  \BibitemOpen
  \bibfield  {author} {\bibinfo {author} {\bibfnamefont {Y.}~\bibnamefont
  {Li}}, \bibinfo {author} {\bibfnamefont {D.}~\bibnamefont {Adroja}}, \bibinfo
  {author} {\bibfnamefont {D.}~\bibnamefont {Voneshen}}, \bibinfo {author}
  {\bibfnamefont {R.~I.}\ \bibnamefont {Bewley}}, \bibinfo {author}
  {\bibfnamefont {Q.}~\bibnamefont {Zhang}}, \bibinfo {author} {\bibfnamefont
  {A.~A.}\ \bibnamefont {Tsirlin}}, \ and\ \bibinfo {author} {\bibfnamefont
  {P.}~\bibnamefont {Gegenwart}},\ }\bibfield  {title} {\enquote {\bibinfo
  {title} {Nearest-neighbor resonating valence bonds in {YbMgGaO$_{4}$}},}\
  }\href {\doibase 10.1038/ncomms15814} {\bibfield  {journal} {\bibinfo
  {journal} {Nat. Commun.}\ }\textbf {\bibinfo {volume} {8}},\ \bibinfo {pages}
  {15814} (\bibinfo {year} {2017}{\natexlab{b}})}\BibitemShut {NoStop}%
\bibitem [{\citenamefont {Han}\ \emph {et~al.}(2012)\citenamefont {Han},
  \citenamefont {Helton}, \citenamefont {Chu}, \citenamefont {Nocera},
  \citenamefont {Rodriguez-Rivera}, \citenamefont {Broholm},\ and\
  \citenamefont {Lee}}]{han2012fractionalized}%
  \BibitemOpen
  \bibfield  {author} {\bibinfo {author} {\bibfnamefont {T.-H.}\ \bibnamefont
  {Han}}, \bibinfo {author} {\bibfnamefont {J.~S.}\ \bibnamefont {Helton}},
  \bibinfo {author} {\bibfnamefont {S.}~\bibnamefont {Chu}}, \bibinfo {author}
  {\bibfnamefont {D.~G.}\ \bibnamefont {Nocera}}, \bibinfo {author}
  {\bibfnamefont {J.~A.}\ \bibnamefont {Rodriguez-Rivera}}, \bibinfo {author}
  {\bibfnamefont {C.}~\bibnamefont {Broholm}}, \ and\ \bibinfo {author}
  {\bibfnamefont {Y.~S.}\ \bibnamefont {Lee}},\ }\bibfield  {title} {\enquote
  {\bibinfo {title} {Fractionalized excitations in the spin-liquid state of a
  kagome-lattice antiferromagnet},}\ }\href@noop {} {\bibfield  {journal}
  {\bibinfo  {journal} {Nature}\ }\textbf {\bibinfo {volume} {492}},\ \bibinfo
  {pages} {406} (\bibinfo {year} {2012})}\BibitemShut {NoStop}%
\bibitem [{\citenamefont {Sakakibara}\ \emph {et~al.}(1994)\citenamefont
  {Sakakibara}, \citenamefont {Mitamura}, \citenamefont {Tayama},\ and\
  \citenamefont {Amitsuka}}]{sakakibara1994faraday}%
  \BibitemOpen
  \bibfield  {author} {\bibinfo {author} {\bibfnamefont {T.}~\bibnamefont
  {Sakakibara}}, \bibinfo {author} {\bibfnamefont {H.}~\bibnamefont
  {Mitamura}}, \bibinfo {author} {\bibfnamefont {T.}~\bibnamefont {Tayama}}, \
  and\ \bibinfo {author} {\bibfnamefont {H.}~\bibnamefont {Amitsuka}},\
  }\bibfield  {title} {\enquote {\bibinfo {title} {Faraday force magnetometer
  for high-sensitivity magnetization measurements at very low temperatures and
  high fields},}\ }\href@noop {} {\bibfield  {journal} {\bibinfo  {journal}
  {Jpn. J. Appl. Phys.}\ }\textbf {\bibinfo {volume} {33}},\ \bibinfo {pages}
  {5067} (\bibinfo {year} {1994})}\BibitemShut {NoStop}%
\bibitem [{\citenamefont {Li}\ \emph {et~al.}(2018{\natexlab{a}})\citenamefont
  {Li}, \citenamefont {Bachus}, \citenamefont {Tokiwa}, \citenamefont
  {Tsirlin},\ and\ \citenamefont {Gegenwart}}]{li2018absence}%
  \BibitemOpen
  \bibfield  {author} {\bibinfo {author} {\bibfnamefont {Y.}~\bibnamefont
  {Li}}, \bibinfo {author} {\bibfnamefont {S.}~\bibnamefont {Bachus}}, \bibinfo
  {author} {\bibfnamefont {Y.}~\bibnamefont {Tokiwa}}, \bibinfo {author}
  {\bibfnamefont {A.~A.}\ \bibnamefont {Tsirlin}}, \ and\ \bibinfo {author}
  {\bibfnamefont {P.}~\bibnamefont {Gegenwart}},\ }\bibfield  {title} {\enquote
  {\bibinfo {title} {Absence of zero-point entropy in a triangular {Ising}
  antiferromagnet},}\ }\href@noop {} {\bibfield  {journal} {\bibinfo  {journal}
  {arXiv preprint arXiv:1804.00696}\ } (\bibinfo {year}
  {2018}{\natexlab{a}})}\BibitemShut {NoStop}%
\bibitem [{\citenamefont {Schneidewind}\ and\ \citenamefont
  {{\v{C}}erm{\'a}k}(2015)}]{schneidewind2015panda}%
  \BibitemOpen
  \bibfield  {author} {\bibinfo {author} {\bibfnamefont {A.}~\bibnamefont
  {Schneidewind}}\ and\ \bibinfo {author} {\bibfnamefont {P.}~\bibnamefont
  {{\v{C}}erm{\'a}k}},\ }\bibfield  {title} {\enquote {\bibinfo {title} {Panda:
  Cold three axes spectrometer},}\ }\href@noop {} {\bibfield  {journal}
  {\bibinfo  {journal} {Journal of large-scale research facilities JLSRF}\
  }\textbf {\bibinfo {volume} {1}},\ \bibinfo {pages} {12} (\bibinfo {year}
  {2015})}\BibitemShut {NoStop}%
\bibitem [{sup()}]{supple}%
  \BibitemOpen
  \href@noop {} {\emph {\bibinfo {title} {See Supplementary material for
  detailed information about experimental procedures, which includes
  Refs.~\cite{shirane2002neutron,xu2013absolute,PhysRevB.65.144421,PhysRevLett.106.187202,zaliznyak2004magnetic}.}}}\BibitemShut
  {Stop}%
\bibitem [{\citenamefont {Fritsch}\ \emph {et~al.}(2017)\citenamefont
  {Fritsch}, \citenamefont {Ross}, \citenamefont {Granroth}, \citenamefont
  {Ehlers}, \citenamefont {Noad}, \citenamefont {Dabkowska},\ and\
  \citenamefont {Gaulin}}]{PhysRevB.96.094414}%
  \BibitemOpen
  \bibfield  {author} {\bibinfo {author} {\bibfnamefont {K.}~\bibnamefont
  {Fritsch}}, \bibinfo {author} {\bibfnamefont {K.~A.}\ \bibnamefont {Ross}},
  \bibinfo {author} {\bibfnamefont {G.~E.}\ \bibnamefont {Granroth}}, \bibinfo
  {author} {\bibfnamefont {G.}~\bibnamefont {Ehlers}}, \bibinfo {author}
  {\bibfnamefont {H.~M.~L.}\ \bibnamefont {Noad}}, \bibinfo {author}
  {\bibfnamefont {H.~A.}\ \bibnamefont {Dabkowska}}, \ and\ \bibinfo {author}
  {\bibfnamefont {B.~D.}\ \bibnamefont {Gaulin}},\ }\bibfield  {title}
  {\enquote {\bibinfo {title} {Quasi-two-dimensional spin correlations in the
  triangular lattice bilayer spin glass {LuCoGaO$_4$}},}\ }\href@noop {}
  {\bibfield  {journal} {\bibinfo  {journal} {Phys. Rev. B}\ }\textbf {\bibinfo
  {volume} {96}},\ \bibinfo {pages} {094414} (\bibinfo {year}
  {2017})}\BibitemShut {NoStop}%
\bibitem [{\citenamefont {Shlyk}\ \emph {et~al.}(2018)\citenamefont {Shlyk},
  \citenamefont {Strobel}, \citenamefont {Farmer}, \citenamefont {De~Long},\
  and\ \citenamefont {Niewa}}]{PhysRevB.97.054426}%
  \BibitemOpen
  \bibfield  {author} {\bibinfo {author} {\bibfnamefont {L.}~\bibnamefont
  {Shlyk}}, \bibinfo {author} {\bibfnamefont {S.}~\bibnamefont {Strobel}},
  \bibinfo {author} {\bibfnamefont {B.}~\bibnamefont {Farmer}}, \bibinfo
  {author} {\bibfnamefont {L.~E.}\ \bibnamefont {De~Long}}, \ and\ \bibinfo
  {author} {\bibfnamefont {R.}~\bibnamefont {Niewa}},\ }\bibfield  {title}
  {\enquote {\bibinfo {title} {Coexistence of ferromagnetism and unconventional
  spin-glass freezing in the site-disordered kagome ferrite
  {SrSn$_2$Fe$_4$O$_{11}$}},}\ }\href@noop {} {\bibfield  {journal} {\bibinfo
  {journal} {Phys. Rev. B}\ }\textbf {\bibinfo {volume} {97}},\ \bibinfo
  {pages} {054426} (\bibinfo {year} {2018})}\BibitemShut {NoStop}%
\bibitem [{\citenamefont {Fisher}(1986)}]{PhysRevLett.56.416}%
  \BibitemOpen
  \bibfield  {author} {\bibinfo {author} {\bibfnamefont {D.~S.}\ \bibnamefont
  {Fisher}},\ }\bibfield  {title} {\enquote {\bibinfo {title} {Scaling and
  critical slowing down in random-field {Ising} systems},}\ }\href@noop {}
  {\bibfield  {journal} {\bibinfo  {journal} {Phys. Rev. Lett.}\ }\textbf
  {\bibinfo {volume} {56}},\ \bibinfo {pages} {416--419} (\bibinfo {year}
  {1986})}\BibitemShut {NoStop}%
\bibitem [{\citenamefont {Vojta}(2003)}]{vojta2003quantum}%
  \BibitemOpen
  \bibfield  {author} {\bibinfo {author} {\bibfnamefont {M.}~\bibnamefont
  {Vojta}},\ }\bibfield  {title} {\enquote {\bibinfo {title} {Quantum phase
  transitions},}\ }\href@noop {} {\bibfield  {journal} {\bibinfo  {journal}
  {Rep. Prog. Phys.}\ }\textbf {\bibinfo {volume} {66}},\ \bibinfo {pages}
  {2069} (\bibinfo {year} {2003})}\BibitemShut {NoStop}%
\bibitem [{\citenamefont {Helton}\ \emph {et~al.}(2010)\citenamefont {Helton},
  \citenamefont {Matan}, \citenamefont {Shores}, \citenamefont {Nytko},
  \citenamefont {Bartlett}, \citenamefont {Qiu}, \citenamefont {Nocera},\ and\
  \citenamefont {Lee}}]{PhysRevLett.104.147201}%
  \BibitemOpen
  \bibfield  {author} {\bibinfo {author} {\bibfnamefont {J.~S.}\ \bibnamefont
  {Helton}}, \bibinfo {author} {\bibfnamefont {K.}~\bibnamefont {Matan}},
  \bibinfo {author} {\bibfnamefont {M.~P.}\ \bibnamefont {Shores}}, \bibinfo
  {author} {\bibfnamefont {E.~A.}\ \bibnamefont {Nytko}}, \bibinfo {author}
  {\bibfnamefont {B.~M.}\ \bibnamefont {Bartlett}}, \bibinfo {author}
  {\bibfnamefont {Y.}~\bibnamefont {Qiu}}, \bibinfo {author} {\bibfnamefont
  {D.~G.}\ \bibnamefont {Nocera}}, \ and\ \bibinfo {author} {\bibfnamefont
  {Y.~S.}\ \bibnamefont {Lee}},\ }\bibfield  {title} {\enquote {\bibinfo
  {title} {Dynamic scaling in the susceptibility of the spin-{$\frac{1}{2}$}
  kagome lattice antiferromagnet {Herbertsmithite}},}\ }\href@noop {}
  {\bibfield  {journal} {\bibinfo  {journal} {Phys. Rev. Lett.}\ }\textbf
  {\bibinfo {volume} {104}},\ \bibinfo {pages} {147201} (\bibinfo {year}
  {2010})}\BibitemShut {NoStop}%
\bibitem [{\citenamefont {Deguchi}\ \emph {et~al.}(2012)\citenamefont
  {Deguchi}, \citenamefont {Matsukawa}, \citenamefont {Sato}, \citenamefont
  {Hattori}, \citenamefont {Ishida}, \citenamefont {Takakura},\ and\
  \citenamefont {Ishimasa}}]{deguchi2012quantum}%
  \BibitemOpen
  \bibfield  {author} {\bibinfo {author} {\bibfnamefont {K.}~\bibnamefont
  {Deguchi}}, \bibinfo {author} {\bibfnamefont {S.}~\bibnamefont {Matsukawa}},
  \bibinfo {author} {\bibfnamefont {N.~K.}\ \bibnamefont {Sato}}, \bibinfo
  {author} {\bibfnamefont {T.}~\bibnamefont {Hattori}}, \bibinfo {author}
  {\bibfnamefont {K.}~\bibnamefont {Ishida}}, \bibinfo {author} {\bibfnamefont
  {H.}~\bibnamefont {Takakura}}, \ and\ \bibinfo {author} {\bibfnamefont
  {T.}~\bibnamefont {Ishimasa}},\ }\bibfield  {title} {\enquote {\bibinfo
  {title} {Quantum critical state in a magnetic quasicrystal},}\ }\href@noop {}
  {\bibfield  {journal} {\bibinfo  {journal} {Nat. Matter.}\ }\textbf {\bibinfo
  {volume} {11}},\ \bibinfo {pages} {1013} (\bibinfo {year}
  {2012})}\BibitemShut {NoStop}%
\bibitem [{\citenamefont {Li}\ \emph {et~al.}(2018{\natexlab{b}})\citenamefont
  {Li}, \citenamefont {Bachus}, \citenamefont {Tokiwa}, \citenamefont
  {Tsirlin},\ and\ \citenamefont {Gegenwart}}]{li2018gapped}%
  \BibitemOpen
  \bibfield  {author} {\bibinfo {author} {\bibfnamefont {Y.}~\bibnamefont
  {Li}}, \bibinfo {author} {\bibfnamefont {S.}~\bibnamefont {Bachus}}, \bibinfo
  {author} {\bibfnamefont {Y.}~\bibnamefont {Tokiwa}}, \bibinfo {author}
  {\bibfnamefont {A.~A.}\ \bibnamefont {Tsirlin}}, \ and\ \bibinfo {author}
  {\bibfnamefont {P.}~\bibnamefont {Gegenwart}},\ }\bibfield  {title} {\enquote
  {\bibinfo {title} {Gapped ground state in the zigzag pseudospin-1/2 quantum
  antiferromagnetic chain compound {PrTiNbO$_6$}},}\ }\href@noop {} {\bibfield
  {journal} {\bibinfo  {journal} {Phys. Rev. B}\ }\textbf {\bibinfo {volume}
  {97}},\ \bibinfo {pages} {184434} (\bibinfo {year}
  {2018}{\natexlab{b}})}\BibitemShut {NoStop}%
\bibitem [{\citenamefont {Yin}\ \emph {et~al.}(2013)\citenamefont {Yin},
  \citenamefont {Xia}, \citenamefont {Takano}, \citenamefont {Sullivan},
  \citenamefont {Li},\ and\ \citenamefont {Sun}}]{PhysRevLett.110.137201}%
  \BibitemOpen
  \bibfield  {author} {\bibinfo {author} {\bibfnamefont {L.}~\bibnamefont
  {Yin}}, \bibinfo {author} {\bibfnamefont {J.~S.}\ \bibnamefont {Xia}},
  \bibinfo {author} {\bibfnamefont {Y.}~\bibnamefont {Takano}}, \bibinfo
  {author} {\bibfnamefont {N.~S.}\ \bibnamefont {Sullivan}}, \bibinfo {author}
  {\bibfnamefont {Q.~J.}\ \bibnamefont {Li}}, \ and\ \bibinfo {author}
  {\bibfnamefont {X.~F.}\ \bibnamefont {Sun}},\ }\bibfield  {title} {\enquote
  {\bibinfo {title} {Low-temperature low-field phases of the pyrochlore quantum
  magnet {Tb$_2$Ti$_2$O$_7$}},}\ }\href@noop {} {\bibfield  {journal} {\bibinfo
   {journal} {Phys. Rev. Lett.}\ }\textbf {\bibinfo {volume} {110}},\ \bibinfo
  {pages} {137201} (\bibinfo {year} {2013})}\BibitemShut {NoStop}%
\bibitem [{\citenamefont {Sears}\ \emph {et~al.}(2017)\citenamefont {Sears},
  \citenamefont {Zhao}, \citenamefont {Xu}, \citenamefont {Lynn},\ and\
  \citenamefont {Kim}}]{PhysRevB.95.180411}%
  \BibitemOpen
  \bibfield  {author} {\bibinfo {author} {\bibfnamefont {J.~A.}\ \bibnamefont
  {Sears}}, \bibinfo {author} {\bibfnamefont {Y.}~\bibnamefont {Zhao}},
  \bibinfo {author} {\bibfnamefont {Z.}~\bibnamefont {Xu}}, \bibinfo {author}
  {\bibfnamefont {J.~W.}\ \bibnamefont {Lynn}}, \ and\ \bibinfo {author}
  {\bibfnamefont {Young-June}\ \bibnamefont {Kim}},\ }\bibfield  {title}
  {\enquote {\bibinfo {title} {Phase diagram of {$\alpha$-RuCl$_3$} in an
  in-plane magnetic field},}\ }\href@noop {} {\bibfield  {journal} {\bibinfo
  {journal} {Phys. Rev. B}\ }\textbf {\bibinfo {volume} {95}},\ \bibinfo
  {pages} {180411} (\bibinfo {year} {2017})}\BibitemShut {NoStop}%
\bibitem [{\citenamefont {T{\'o}th}\ \emph {et~al.}(2017)\citenamefont
  {T{\'o}th}, \citenamefont {Rolfs}, \citenamefont {Wildes},\ and\
  \citenamefont {R{\"u}egg}}]{toth2017strong}%
  \BibitemOpen
  \bibfield  {author} {\bibinfo {author} {\bibfnamefont {S.}~\bibnamefont
  {T{\'o}th}}, \bibinfo {author} {\bibfnamefont {K.}~\bibnamefont {Rolfs}},
  \bibinfo {author} {\bibfnamefont {A.~R.}\ \bibnamefont {Wildes}}, \ and\
  \bibinfo {author} {\bibfnamefont {C.}~\bibnamefont {R{\"u}egg}},\ }\bibfield
  {title} {\enquote {\bibinfo {title} {Strong exchange anisotropy in
  {YbMgGaO$_4$} from polarized neutron diffraction},}\ }\href@noop {}
  {\bibfield  {journal} {\bibinfo  {journal} {arXiv preprint arXiv:1705.05699}\
  } (\bibinfo {year} {2017})}\BibitemShut {NoStop}%
\bibitem [{\citenamefont {Rotter}\ and\ \citenamefont
  {Boothroyd}(2009)}]{PhysRevB.79.140405}%
  \BibitemOpen
  \bibfield  {author} {\bibinfo {author} {\bibfnamefont {M.}~\bibnamefont
  {Rotter}}\ and\ \bibinfo {author} {\bibfnamefont {A.~T.}\ \bibnamefont
  {Boothroyd}},\ }\bibfield  {title} {\enquote {\bibinfo {title} {Going beyond
  the dipole approximation to improve the refinement of magnetic structures by
  neutron diffraction},}\ }\href@noop {} {\bibfield  {journal} {\bibinfo
  {journal} {Phys. Rev. B}\ }\textbf {\bibinfo {volume} {79}},\ \bibinfo
  {pages} {140405} (\bibinfo {year} {2009})}\BibitemShut {NoStop}%
\bibitem [{Note2()}]{Note2}%
  \BibitemOpen
  \bibinfo {note} {The $S_{b1}+f_2S_{b2}+f_3S_{b3}+f_4S_{b4}$ fits yield
  $|f_4|$ $<$ 0.04, and the least-$R_p$ decreases by less than 0.02, compared
  to the $S_{b1}+f_2S_{b2}+f_3S_{b3}$ fits}\BibitemShut {NoStop}%
\bibitem [{\citenamefont {Lake}\ \emph {et~al.}(2005)\citenamefont {Lake},
  \citenamefont {Tennant}, \citenamefont {Frost},\ and\ \citenamefont
  {Nagler}}]{lake2005quantum}%
  \BibitemOpen
  \bibfield  {author} {\bibinfo {author} {\bibfnamefont {B.}~\bibnamefont
  {Lake}}, \bibinfo {author} {\bibfnamefont {D.~A.}\ \bibnamefont {Tennant}},
  \bibinfo {author} {\bibfnamefont {C.~D.}\ \bibnamefont {Frost}}, \ and\
  \bibinfo {author} {\bibfnamefont {S.~E.}\ \bibnamefont {Nagler}},\ }\bibfield
   {title} {\enquote {\bibinfo {title} {Quantum criticality and universal
  scaling of a quantum antiferromagnet},}\ }\href@noop {} {\bibfield  {journal}
  {\bibinfo  {journal} {Nat. Mater.}\ }\textbf {\bibinfo {volume} {4}},\
  \bibinfo {pages} {329} (\bibinfo {year} {2005})}\BibitemShut {NoStop}%
\bibitem [{\citenamefont {Lake}\ \emph {et~al.}(2013)\citenamefont {Lake},
  \citenamefont {Tennant}, \citenamefont {Caux}, \citenamefont {Barthel},
  \citenamefont {Schollw\"ock}, \citenamefont {Nagler},\ and\ \citenamefont
  {Frost}}]{PhysRevLett.111.137205}%
  \BibitemOpen
  \bibfield  {author} {\bibinfo {author} {\bibfnamefont {B.}~\bibnamefont
  {Lake}}, \bibinfo {author} {\bibfnamefont {D.~A.}\ \bibnamefont {Tennant}},
  \bibinfo {author} {\bibfnamefont {J.-S.}\ \bibnamefont {Caux}}, \bibinfo
  {author} {\bibfnamefont {T.}~\bibnamefont {Barthel}}, \bibinfo {author}
  {\bibfnamefont {U.}~\bibnamefont {Schollw\"ock}}, \bibinfo {author}
  {\bibfnamefont {S.~E.}\ \bibnamefont {Nagler}}, \ and\ \bibinfo {author}
  {\bibfnamefont {C.~D.}\ \bibnamefont {Frost}},\ }\bibfield  {title} {\enquote
  {\bibinfo {title} {Multispinon continua at zero and finite temperature in a
  near-ideal {Heisenberg} chain},}\ }\href {\doibase
  10.1103/PhysRevLett.111.137205} {\bibfield  {journal} {\bibinfo  {journal}
  {Phys. Rev. Lett.}\ }\textbf {\bibinfo {volume} {111}},\ \bibinfo {pages}
  {137205} (\bibinfo {year} {2013})}\BibitemShut {NoStop}%
\bibitem [{\citenamefont {Banerjee}\ \emph {et~al.}(2017)\citenamefont
  {Banerjee}, \citenamefont {Yan}, \citenamefont {Knolle}, \citenamefont
  {Bridges}, \citenamefont {Stone}, \citenamefont {Lumsden}, \citenamefont
  {Mandrus}, \citenamefont {Tennant}, \citenamefont {Moessner},\ and\
  \citenamefont {Nagler}}]{banerjee2017neutron}%
  \BibitemOpen
  \bibfield  {author} {\bibinfo {author} {\bibfnamefont {A.}~\bibnamefont
  {Banerjee}}, \bibinfo {author} {\bibfnamefont {J.}~\bibnamefont {Yan}},
  \bibinfo {author} {\bibfnamefont {J.}~\bibnamefont {Knolle}}, \bibinfo
  {author} {\bibfnamefont {C.~A.}\ \bibnamefont {Bridges}}, \bibinfo {author}
  {\bibfnamefont {M.~B.}\ \bibnamefont {Stone}}, \bibinfo {author}
  {\bibfnamefont {M.~D.}\ \bibnamefont {Lumsden}}, \bibinfo {author}
  {\bibfnamefont {D.~G.}\ \bibnamefont {Mandrus}}, \bibinfo {author}
  {\bibfnamefont {D.~A.}\ \bibnamefont {Tennant}}, \bibinfo {author}
  {\bibfnamefont {R.}~\bibnamefont {Moessner}}, \ and\ \bibinfo {author}
  {\bibfnamefont {S.~E.}\ \bibnamefont {Nagler}},\ }\bibfield  {title}
  {\enquote {\bibinfo {title} {Neutron scattering in the proximate quantum spin
  liquid {$\alpha$-RuCl$_3$}},}\ }\href@noop {} {\bibfield  {journal} {\bibinfo
   {journal} {Science}\ }\textbf {\bibinfo {volume} {356}},\ \bibinfo {pages}
  {1055--1059} (\bibinfo {year} {2017})}\BibitemShut {NoStop}%
\bibitem [{\citenamefont {Do}\ \emph {et~al.}(2017)\citenamefont {Do},
  \citenamefont {Park}, \citenamefont {Yoshitake}, \citenamefont {Nasu},
  \citenamefont {Motome}, \citenamefont {Kwon}, \citenamefont {Adroja},
  \citenamefont {Voneshen}, \citenamefont {Kim}, \citenamefont {Jang},
  \citenamefont {Park}, \citenamefont {Choi},\ and\ \citenamefont
  {Ji}}]{do2017majorana}%
  \BibitemOpen
  \bibfield  {author} {\bibinfo {author} {\bibfnamefont {S.-H.}\ \bibnamefont
  {Do}}, \bibinfo {author} {\bibfnamefont {S.-Y.}\ \bibnamefont {Park}},
  \bibinfo {author} {\bibfnamefont {J.}~\bibnamefont {Yoshitake}}, \bibinfo
  {author} {\bibfnamefont {J.}~\bibnamefont {Nasu}}, \bibinfo {author}
  {\bibfnamefont {Y.}~\bibnamefont {Motome}}, \bibinfo {author} {\bibfnamefont
  {Y.~S.}\ \bibnamefont {Kwon}}, \bibinfo {author} {\bibfnamefont {D.~T.}\
  \bibnamefont {Adroja}}, \bibinfo {author} {\bibfnamefont {D.~J.}\
  \bibnamefont {Voneshen}}, \bibinfo {author} {\bibfnamefont {K.}~\bibnamefont
  {Kim}}, \bibinfo {author} {\bibfnamefont {T.-H.}\ \bibnamefont {Jang}},
  \bibinfo {author} {\bibfnamefont {J.-H.}\ \bibnamefont {Park}}, \bibinfo
  {author} {\bibfnamefont {K.-Y.}\ \bibnamefont {Choi}}, \ and\ \bibinfo
  {author} {\bibfnamefont {S.}~\bibnamefont {Ji}},\ }\bibfield  {title}
  {\enquote {\bibinfo {title} {{Majorana} fermions in the {Kitaev} quantum spin
  system {$\alpha$-RuCl$_3$}},}\ }\href@noop {} {\bibfield  {journal} {\bibinfo
   {journal} {Nat. Phys.}\ }\textbf {\bibinfo {volume} {13}},\ \bibinfo {pages}
  {1079} (\bibinfo {year} {2017})}\BibitemShut {NoStop}%
\bibitem [{\citenamefont {Shirane}\ \emph {et~al.}(2002)\citenamefont
  {Shirane}, \citenamefont {Shapiro},\ and\ \citenamefont
  {Tranquada}}]{shirane2002neutron}%
  \BibitemOpen
  \bibfield  {author} {\bibinfo {author} {\bibfnamefont {G.}~\bibnamefont
  {Shirane}}, \bibinfo {author} {\bibfnamefont {S.~M.}\ \bibnamefont
  {Shapiro}}, \ and\ \bibinfo {author} {\bibfnamefont {J.~M.}\ \bibnamefont
  {Tranquada}},\ }\href@noop {} {\emph {\bibinfo {title} {Neutron scattering
  with a triple-axis spectrometer: basic techniques}}}\ (\bibinfo  {publisher}
  {Cambridge University Press},\ \bibinfo {year} {2002})\BibitemShut {NoStop}%
\bibitem [{\citenamefont {Xu}\ \emph {et~al.}(2013)\citenamefont {Xu},
  \citenamefont {Xu},\ and\ \citenamefont {Tranquada}}]{xu2013absolute}%
  \BibitemOpen
  \bibfield  {author} {\bibinfo {author} {\bibfnamefont {G.}~\bibnamefont
  {Xu}}, \bibinfo {author} {\bibfnamefont {Z.}~\bibnamefont {Xu}}, \ and\
  \bibinfo {author} {\bibfnamefont {J.~M.}\ \bibnamefont {Tranquada}},\
  }\bibfield  {title} {\enquote {\bibinfo {title} {Absolute cross-section
  normalization of magnetic neutron scattering data},}\ }\href@noop {}
  {\bibfield  {journal} {\bibinfo  {journal} {Rev. Sci. Instrum.}\ }\textbf
  {\bibinfo {volume} {84}},\ \bibinfo {pages} {083906} (\bibinfo {year}
  {2013})}\BibitemShut {NoStop}%
\bibitem [{\citenamefont {Kadowaki}\ \emph {et~al.}(2002)\citenamefont
  {Kadowaki}, \citenamefont {Ishii}, \citenamefont {Matsuhira},\ and\
  \citenamefont {Hinatsu}}]{PhysRevB.65.144421}%
  \BibitemOpen
  \bibfield  {author} {\bibinfo {author} {\bibfnamefont {H.}~\bibnamefont
  {Kadowaki}}, \bibinfo {author} {\bibfnamefont {Y.}~\bibnamefont {Ishii}},
  \bibinfo {author} {\bibfnamefont {K.}~\bibnamefont {Matsuhira}}, \ and\
  \bibinfo {author} {\bibfnamefont {Y.}~\bibnamefont {Hinatsu}},\ }\bibfield
  {title} {\enquote {\bibinfo {title} {Neutron scattering study of dipolar spin
  ice {Ho$_2$Sn$_2$O$_7$}: {Frustrated} pyrochlore magnet},}\ }\href {\doibase
  10.1103/PhysRevB.65.144421} {\bibfield  {journal} {\bibinfo  {journal} {Phys.
  Rev. B}\ }\textbf {\bibinfo {volume} {65}},\ \bibinfo {pages} {144421}
  (\bibinfo {year} {2002})}\BibitemShut {NoStop}%
\bibitem [{\citenamefont {Thompson}\ \emph {et~al.}(2011)\citenamefont
  {Thompson}, \citenamefont {McClarty}, \citenamefont {R{\o}nnow},
  \citenamefont {Regnault}, \citenamefont {Sorge},\ and\ \citenamefont
  {Gingras}}]{PhysRevLett.106.187202}%
  \BibitemOpen
  \bibfield  {author} {\bibinfo {author} {\bibfnamefont {J.~D.}\ \bibnamefont
  {Thompson}}, \bibinfo {author} {\bibfnamefont {P.~A.}\ \bibnamefont
  {McClarty}}, \bibinfo {author} {\bibfnamefont {H.~M.}\ \bibnamefont
  {R{\o}nnow}}, \bibinfo {author} {\bibfnamefont {L.~P.}\ \bibnamefont
  {Regnault}}, \bibinfo {author} {\bibfnamefont {A.}~\bibnamefont {Sorge}}, \
  and\ \bibinfo {author} {\bibfnamefont {M.~J.~P.}\ \bibnamefont {Gingras}},\
  }\bibfield  {title} {\enquote {\bibinfo {title} {Rods of neutron scattering
  intensity in {Yb$_2$Ti$_2$O$_7$}: {Compelling} evidence for significant
  anisotropic exchange in a magnetic pyrochlore oxide},}\ }\href {\doibase
  10.1103/PhysRevLett.106.187202} {\bibfield  {journal} {\bibinfo  {journal}
  {Phys. Rev. Lett.}\ }\textbf {\bibinfo {volume} {106}},\ \bibinfo {pages}
  {187202} (\bibinfo {year} {2011})}\BibitemShut {NoStop}%
\bibitem [{\citenamefont {Zaliznyak}\ and\ \citenamefont
  {Lee}(2004)}]{zaliznyak2004magnetic}%
  \BibitemOpen
  \bibfield  {author} {\bibinfo {author} {\bibfnamefont {I.~A.}\ \bibnamefont
  {Zaliznyak}}\ and\ \bibinfo {author} {\bibfnamefont {S.-H.}\ \bibnamefont
  {Lee}},\ }\href@noop {} {\emph {\bibinfo {title} {Magnetic neutron
  scattering}}},\ \bibinfo {type} {Tech. Rep.}\ (\bibinfo  {institution}
  {BROOKHAVEN NATIONAL LABORATORY (US)},\ \bibinfo {year} {2004})\BibitemShut
  {NoStop}%
\end{thebibliography}%

\end{document}